# Theory of Edge Effects and Conductance for Applications in Graphene-based Nanoantennas

**Tomer Berghaus**[1]*, **Touvia Miloh**[1], **Oded Gottlieb**[2] and **Gregory Ya. Slepyan**[3]*

[1] School of Mechanical Engineering, Tel Aviv University, Tel Aviv 69978, Israel
[2] Faculty of Mechanical Engineering, Technion - Israel Institute of Technology, Haifa, 3200003, Israel
[3] School of Electrical Engineering, Tel Aviv University, Tel Aviv 69978, Israel
* Correspondence: berghaus@gmail.com (T.B); gregory_slepyan@yahoo.com (G.S.)

**Abstract**: In this paper, we develop a theory of edge effects in graphene for its applications to nanoantennas in the THz, infrared, and visible frequency ranges. Its characteristic feature is self-consistence reached due the formulation in terms of dynamical conductance instead of ordinary used surface conductivity. The physical model of edge effects is based on using the concept of Dirac fermions. The surface conductance is considered as a general susceptibility and is calculated via the Kubo approach. In contrast with earlier models, the surface conductance becomes non-homogeneous and non-local. The spatial behavior of the surface conductance depends on the length of the sheet and the electrochemical potential. Results of numerical simulations are presented for lengths in the range of 2.1 - 800nm and electrochemical potentials ranging between $0.1 - 1.0$ eV. It is shown that if the length exceeded 800 nm, our model agrees with the classical Drude conductivity model with a relatively high degree of accuracy. For rather short lengths, the conductance usually exhibits spatial oscillations, which absent in conductivity and strongly affect the properties of graphene based antennas. The period and amplitude of such spatial oscillations, strongly depend on the electrochemical potential. The new theory opens the way for realizing electrically controlled nanoantennas by changing the electrochemical potential may of the gate voltage. The obtained results may be applicable for the design of carbon based nanodevices in modern quantum technologies.

1. **Introduction.**

Innovative electromagnetic nano-antennas, which generally function at wide range of frequencies (from THz until video frequencies), play a vital role in the emerging field of photonics and plasmonics [1-6]. These antennas can be used as promising tools for transforming near-field light into far-field and vice versa. Their excellent capabilities are usually utilized in a wide scope of applications. Among them, are traditional applications for basic elements in electronics, high-speed communications, informatics, and quantum computing [7] (in particular quantum nano-mechanical qubit on carbon nanotube [8]). As far as commercial applications of carbon-based nanostructures [9], one can list some novel applications such as: i) high-resolved spectroscopy [10]; ii) high-speed communication [11]; iii) light emission and detection; iv) identification of biomolecules and medical diagnostic applications [12,13]. The design of the devices mentioned above, requires taking into account edge effects and their correct physical description. It should cover a wide frequency range from microwave until optical. These recent innovations have stimulated huge interest in the theory of Nano-antennas. Microwave standards become invalid starting from the THz region, as the size of the antennas is miniaturized to micrometer scale [6]. Therefore, the common model of perfect electric conductor, which is widely used in microwaves, is not suitable in the range from THz to optical frequencies. By manipulating the different types of finite-size (edge) effects [14-16], the conventional antenna configurations can be also used in the range from microwave to optical frequencies [1-6]. However, antenna devices which operate in the terahertz range (0.1–10 THz), have some fundamental limitations on their applicability in practical devices [6]. Recent studies have pointed out that graphene is one of the best candidates for overcoming the limitations mentioned above [6].

A nanoantenna in its theoretical analysis, is generally considered as a system terminated by a well-defined edge. The edge geometry is one of the widely used idealizations in modern physics. For example, in all branches of classical mechanics and physics (e.g., elasticity, hydrodynamics, acoustics, electrodynamics), the edge has a configurational character. It corresponds to a point or contour at the surface of the body on which the normal vector is usually undefined (half-plane, wedge, cone, etc.). Taking the edge position into consideration, allows us to simplify the boundary conditions and employ certain analytical techniques such as separation of variables, using for example special orthogonal coordinate systems [17] or the Wiener-Hopf method [18]. Such analytical solutions are especially attractive for the qualitative analysis of novel types of problems involving interacting fields of different physical origin (for example, plasmonics and optoelectrofluidics [19]). However, such idealization may lead to the loss of uniqueness of the problem statement. The reason for it is that the placement of an arbitrary point singularity at the edge, generally does not violate the governing wave equation as well as the boundary and radiation conditions. The analytic solutions thus obtained, are correct from the mathematical point of view. However, they may turn to be physically incorrect because due to lack of uniqueness, they may correspond to another source of the field. In order to obtain the corresponding of unique solution, one must enforce the condition of finite (bounded) energy over an arbitrary area (finite extend) of space (Meixner condition) [20]. Such an approach usually leads to the creation of field singularities at the vicinity of the edge. Note, that the edge-like configuration does not change the constitutive properties of the medium (for example, such as a perfect conductor or a perfect insulator in classical electrodynamics).

The extensive recent progress in nanotechnologies, lead to the discovery of novel artificial types of condensed matter, such as graphene [21], carbon nanotubes (CNTs) [22], Weyl and Dirac semimetals [23] and topological insulators [24,25]. Consequently, a great number of fundamental physical problems were reconsidered and in particular, but nevertheless most of them failed to consider the important edge concept. The edge for the different types of nanostructures corresponds to the spatial area in the vicinity of the sample boundary which size is rather large compared with the inter-atomic distance. However, the mechanism of electronic transport at the edge is dramatically different from the corresponding properties of the bulk region. One of the reasons for this disparity is the existence of new types of quantum states strongly confined to the vicinity of the edge, which are able to produce novel physical properties such as topological order (so called, edge states [16,21,25]). Such transport mechanisms manifest themselves in the special optical and optomechanical properties of different types of metamaterials (e.g., carbon based nanostructures).

The electrical properties of macroscopic structures may be described both in terms of conductivity and conductance [16], which are equivalent and coupled via the simple constant coefficient defined by the size values of the sample. As it was noted in [16], the conductivity for nanomaterials (including graphene) is a well-defined value only when the sample is enough large. It is clear from intuitive point of view, that the electric current becomes homogeneous and insensitive to boundaries of the sample. When the sample's size is reduced, the current becomes non-homogeneous and non-local with respect to the field variations in the material. Then, the concept of conductivity loses its meaning. The behavior of the electrons in nanoantennas becomes sensitive to the feeding lines, detectors, modulators and the edges due to the quantum-mechanical interference. As we will show, exactly optical conductance will be suitable parameter for formulation of the effective boundary conditions for electromagnetic field in nanoantennas. In contrast with conductivity, it is non-homogeneous and not coupled via simple relation with optical conductivity (which is homogeneous value). It is described by the Kubo approach instead of Boltzmann's transport equation.

The edge areas play the main role in forming the emission in classical radio-frequency antennas [26] and optical nanoantennas [27]. One of the promising types of nanoantennas that operate in the THz and optical frequency range, are based on carbon-based nanostructures (graphene, CNTs) [28-31]. The physical mechanism of their radiation, is based on the existence of a strongly retarded surface Plasmon predicted in [28, 32], experimentally observed in [33] and used for interpretation of the intriguing measurements of the THz conductivity peak [34]. These works use different models for the charge transport [28, 32, 34, 35], but all of them are not self-consistent. In another words, the boundary-value problem for the Maxwell equations is formulated for the real geometry of the object, while the value of conductivity is introduced as a phenomenological parameter which is determined by using a model corresponding to an infinitely large structure. Of course, such an approach appears to be attractive since it allows to reach, simplification due to the separation of the EM-field and charge transport equations. Such an approach keeps the self-consistent requirement for planar structures with the Fresnel transmission-reflection condition of EM-fields (graphene films, semitransparent mirrors, etc.) [37-40]. However, it may not be adequate for the detailed description of EM-field scattering and antenna emission in the general case. One can clearly expect that the error of such a non-consistent approach may be ignored as being very small, providing the size of the system is rather large. However, it means, that the effect of the Fresnel transmission-reflection dominates the field forming, whereas the antenna efficiency decreases. Nevertheless, it is impossible to say a priori what is the validity bound of such a simplification.

This problem becomes especially relevant for graphene-based structures, because of the corresponding large geometrical size disparity for applications in different graphene devices. The advanced graphene synthesis methods make it possible to grow graphene samples from 2.1nm to few centimeters [6,41-46]. The detail description of optical graphene conductivity requires a self-consistent analysis. Such a self-consistent approach determines the field acting on electrons produced over their motion, by taking into consideration the finite-size configuration and using an appropriate microscopic model for the edge. The formulation of such a self-consistent analysis is one of the main contributions of this paper. One of the important results of this paper, is analyzing the effect of the edges (shape of finite extend) and determine the corresponding error when self-consistency is not prevailed. It is important to note; that edge effects do not only add up to the quantitative differences in the value of conductivity. Edge effects are also able to dramatically affect the special physical mechanism of charge transport. It makes required the formulation of the theory in terms of optical conductance, which is important for applications in antenna design.

The paper is organized as follows: models of the edge of a graphene ribbon based on the Dirac-fermion concept, are first discussed in Section 2. Thereafter, based on using the Kubo approach and the concept of general susceptibility [47], we analytically obtain an expression for the surface conductance of a terminated (finite) graphene sheet. Results of numerical simulations together with a discussion are presented in Section 3. Conclusion and outlook are finally given in Section 4.

## 2. Optical conductance of a terminated graphene sheet.

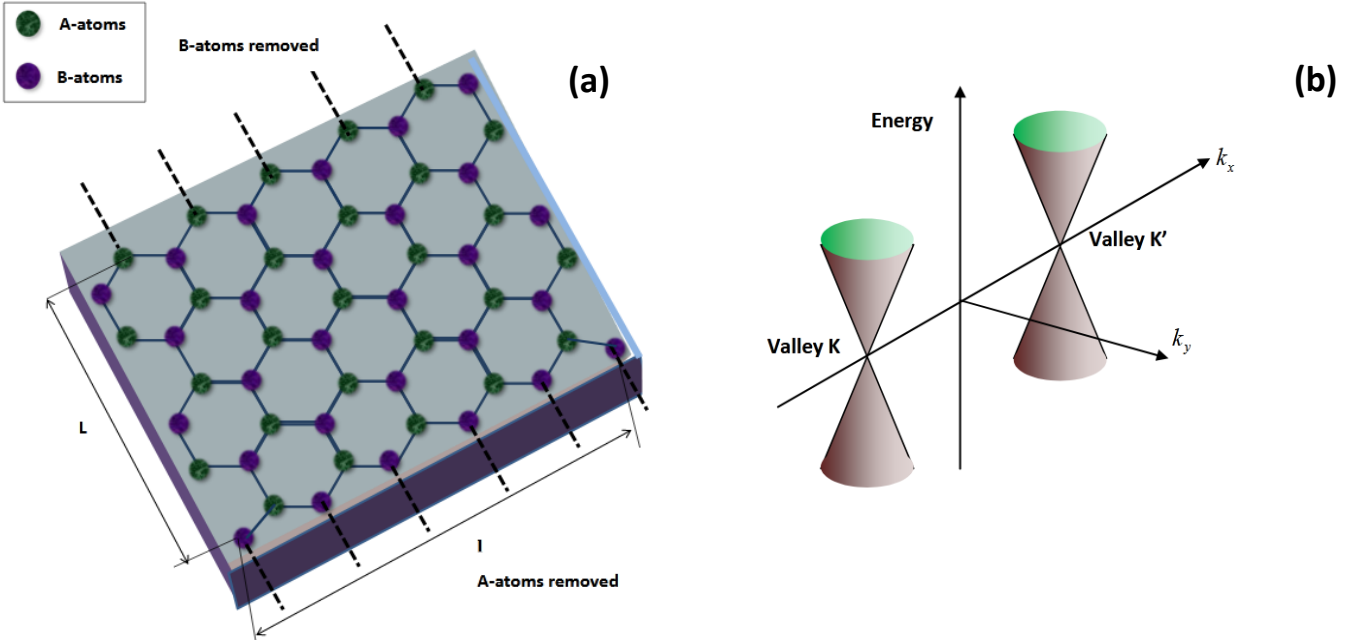

**Figure 1**. Problem statement: **(a)** Geometry of a graphene sheet with zigzag edge configuration; **(b)** Fermi-Dirac cones for the model of pseudospin dispersion.

*2.1 Kubo approach for optical conductance of graphene*

In the following we will use the Kubo approach for conductance calculation as a universal technique which couples the generalized forces of arbitrary physical origin with the responses of correspondent origin via general susceptibilities [47]. Towards this goal it is convenient to define the forces as tangential components of electric field and responses as components of current densities. Thus, the general susceptibilities are defined by a $2\times 2$ matrix where

$$\bar{j}_{a,\omega}(\mathbf{x}) = -\sum_{b=1,2}\int K_{ab}(\omega;\mathbf{x},\mathbf{x}')E_{b,\omega}(\mathbf{x}')d\mathbf{x}' \quad (1)$$

and

$$K_{ab}(\omega;\mathbf{x},\mathbf{x}') = \frac{e^2\omega}{\hbar}\int_0^\infty e^{i\omega t}\left\langle \hat{x}_a(t,\mathbf{x})\hat{x}_b(0,\mathbf{x}') - \hat{x}_b(0,\mathbf{x}')\hat{x}_a(t,\mathbf{x})\right\rangle dt \quad (2)$$

Here $\bar{j}_a(\mathbf{x}) = \langle \hat{j}_a(\mathbf{x})\rangle$, $\hat{j}_a(t,\mathbf{x}) = i\omega e \hat{x}_a(t,\mathbf{x})$, $\langle \hat{j}_a(\mathbf{x})\rangle$ represents the observable current density, $\hat{x}_a(t,\mathbf{x})$ is an operator of charge displacement per unit area. The general susceptibility tensor $K_{ab}(\omega;\mathbf{x},\mathbf{x}')$ depends on the geometry of the sample as well as the electronic properties of the medium. Therefore, exactly it has a physical meaning of non-local optical conductance (in the units $e^2/\hbar$) similar to the dc conductance of graphene in [16].

The next step is the transformation of Equation (2) to a form that is convenient for applications in graphene electrodynamics. Thus one can write

$$K_{ab}(\omega;\mathbf{x},\mathbf{x}') = -\frac{ie}{\hbar}\int_0^\infty e^{i\omega t}\left(\langle \hat{A}^{ab}\rangle - \langle \hat{B}^{ab}\rangle\right)dt \tag{3}$$

where $\langle \hat{A}^{ab}\rangle = \langle \hat{j}_a(t,\mathbf{x})\hat{x}_b(0,\mathbf{x}')\rangle$, $\langle \hat{B}^{ab}\rangle = \langle \hat{x}_b(0,\mathbf{x}')\hat{j}_a(t,\mathbf{x})\rangle$ and

$$\langle \hat{A}^{ab}\rangle = Tr(\hat{A}^{ab}\hat{\rho}_0) = \sum_s \left(\hat{A}^{ab}\right)_{ss}\rho_{0ss} \tag{4}$$

where $\hat{\rho}_0$ is the density matrix of free motion.

The diagonal matrix elements of the operator $\hat{A}^{ab}$ are defined by

$$\left(\hat{A}^{ab}\right)_{ss} = \sum_{s'}\left(\hat{j}_a(t,\mathbf{x})\right)_{ss'}\left(\hat{x}_b(0,\mathbf{x}')\right)_{s's} \tag{5}$$

and following Equation (4) one gets

$$\langle \hat{A}^{ab}\rangle = \sum_s\sum_{s'}\left(\hat{j}_a(t,\mathbf{x})\right)_{ss'}\cdot\left(\hat{x}_b(0,\mathbf{x}')\right)_{s's}\rho_{0ss} \tag{6}$$

Here $s = \{p,k_y\}$ denotes the combinative discrete-continuous index ($p$ is the discrete number of the state and $k_y$ is its continuous wave-number over the *y*-axis).

The temporal behavior in the linear approximation can be expressed as $\left(\hat{j}_a(t,\mathbf{x})\right)_{ss'} = \left(\hat{j}_a(0,\mathbf{x})\right)_{ss'}e^{-i(\varepsilon_s - \varepsilon_{s'})t/\hbar}$ and as a result we obtain

$$\langle \hat{A}^{ab}\rangle = \sum_s\sum_{s'}\left(\hat{j}_a(0,\mathbf{x})\right)_{ss'}\cdot\left(\hat{x}_b(0,\mathbf{x}')\right)_{s's}\rho_{0ss}e^{-\frac{i}{\hbar}(\varepsilon_s - \varepsilon_{s'})t} \tag{7}$$

A similar relation may be also obtained for $\hat{B}^{ab}$, namely

$$\langle \hat{B}^{ab}\rangle = \sum_s\sum_{s'}\left(\hat{x}_b(0,\mathbf{x}')\right)_{ss'}\cdot\left(\hat{j}_a(0,\mathbf{x})\right)_{s's}\rho_{0ss}e^{\frac{i}{\hbar}(\varepsilon_s - \varepsilon_{s'})t} \tag{8}$$

Combining Equations (7) and (8) with Equation (3) lead to

$$K_{ab}(\omega;\mathbf{x},\mathbf{x}') = \frac{i}{4\hbar\omega}\sum_s\sum_{s'}\cdot$$
$$\int_0^\infty e^{i\omega t}\left\{\left(\hat{j}_a(0,\mathbf{x})\right)_{ss'}\left(\hat{x}_b(0,\mathbf{x}')\right)_{s's}e^{-\frac{i}{\hbar}(\varepsilon_s - \varepsilon_{s'})t} - \left(\hat{x}_b(0,\mathbf{x}')\right)_{ss'}\left(\hat{j}_a(0,\mathbf{x})\right)_{s's}e^{\frac{i}{\hbar}(\varepsilon_s - \varepsilon_{s'})t}\right\}\rho_{0ss}dt$$
$$\tag{9}$$

For shortness, we will omit the initial value at time *t*=0 in the matrix elements of $\left(\left(\hat{j}_a(0,\mathbf{x})\right)_{ss'} \to \left(\hat{j}_a(\mathbf{x})\right)_{ss'}\right.$, etc.). Using elementary integration yields

$$K_{ab}(\omega;\mathbf{x},\mathbf{x}') = -\frac{ie}{\hbar}\sum_s\sum_{s'}\left\{\frac{\left(\hat{j}_a(\mathbf{x})\right)_{ss'}\left(\hat{x}_b(\mathbf{x}')\right)_{s's}}{\left(\hbar\omega-(\varepsilon_s-\varepsilon_{s'})\right)} - \frac{\left(\hat{x}_b(\mathbf{x}')\right)_{ss'}\left(\hat{j}_a(\mathbf{x})\right)_{s's}}{\left(\hbar\omega+(\varepsilon_s-\varepsilon_{s'})\right)}\right\}\rho_{0ss} \qquad (10)$$

Introducing the following change $s \to s', s' \to s$ in the second term of Equation (10) and using the relation $\left(\hat{j}_a(\mathbf{x})\right)_{ss'} = ie(\varepsilon_{s'}-\varepsilon_s)\left(\hat{x}_a(\mathbf{x})\right)_{ss'}/\hbar$, the latter can be written as

$$K_{ab}(\omega;\mathbf{x},\mathbf{x}') = -i\hbar\sum_s\sum_{s'}\frac{\left(\hat{j}_a(\mathbf{x})\right)_{ss'}\left(\hat{j}_b(\mathbf{x}')\right)_{s's}}{\left(\hbar\omega-(\varepsilon_s-\varepsilon_{s'})\right)}(\rho_{0ss}-\rho_{0s's'})\frac{1}{(\varepsilon_s-\varepsilon_{s'})} \qquad (11)$$

where $\rho_{0ss} = f(\varepsilon) = 1/\left(e^{(\varepsilon-\mu)/k_BT}+1\right)$ denotes the Fermi-distribution and $\mu$ is the electrochemical potential. The electrochemical potential is defined here by the concentration of electrons/holes according to the following relation; $n_0 = \pi^{-1}(\mu/\hbar v_F)^2$ [37-40]. This potential vanishes for a perfectly clean graphene at zero temperature [16] and may be effectively controlled via the gate voltage [37-40].

Because of the Hermittivity nature of the current operator, we have $\left(\hat{j}_b(\mathbf{x})\right)_{s's} = \left(\hat{j}_b^*(\mathbf{x})\right)_{ss'}$ where the upper asterisk denotes complex conjugate. Therefore, one can also write Equation (11) in the . compact form

$$K_{ab}(\omega;\mathbf{x},\mathbf{x}') = -\frac{i}{\omega}\left(\Pi_{ab}(\omega)-\Pi_{ab}(0)\right) \qquad (12)$$

where

$$\Pi_{ab}(\omega) = \sum_s\sum_{s'}\frac{\left(\hat{j}_a(\mathbf{x})\right)_{ss'}\left(\hat{j}_b(\mathbf{x}')\right)_{s's}}{\left(\hbar\omega-(\varepsilon_s-\varepsilon_{s'})\right)}\left(f(\varepsilon_s)-f(\varepsilon_{s'})\right). \qquad (13)$$

## 2.2. Zigzag edges

In this Subsection we will apply the general result to the edge of a zigzag type. The electronic properties of the ribbon are described by the model of Dirac fermions corresponding to a tight-binding model on a two-dimensional honeycomb lattice [16,25,48]. We will use the zigzag-type of boundary conditions for Dirac fermions on a terminated (finite) lattice derived in [16,25,48], since it was demonstrated that this type of boundary condition can be generally applied to a terminated honeycomb lattice in the case of electron-hole symmetry [25]).
The A and B atoms are coupled in every spinor modes. From physical point of view, the interpretation of electronic transport may be considered as a motion from A atoms to B ones and vice versa, while A-A and B-B motions are forbidden. The electron transport for a zigzag edge in valleys K and K' is independent and will be considered separately. The wave function is expressed as a linear combination of the eigen 4D spinors

The pseudo-spinor modes for valley K have the following form;

$$\Psi_{Ks}(\mathbf{x},t) = \begin{pmatrix} \Psi_{A,s}(\mathbf{x},t) \\ \Psi_{B,s}(\mathbf{x},t) \\ 0 \\ 0 \end{pmatrix} = \begin{pmatrix} u_s(x) \\ v_s(x) \\ 0 \\ 0 \end{pmatrix} e^{i(k_y y - \omega_s t)} \quad (14a)$$

$$\Psi_{K's}(\mathbf{x},t) = \begin{pmatrix} 0 \\ 0 \\ -\tilde{\Psi}_{A,s}(\mathbf{x},t) \\ -\tilde{\Psi}_{B,s}(\mathbf{x},t) \end{pmatrix} = \begin{pmatrix} 0 \\ 0 \\ -\tilde{u}_s(x) \\ -\tilde{v}_s(x) \end{pmatrix} e^{i(k_y y - \omega_s t)} \quad (14b)$$

The functions $u(x), v(x)$ in the valley K (and in K' respectively) take the following form for bulk and edge states

$$\left. \begin{aligned} u_s(x) &= u_s^{Bulk}(x) = i\sqrt{\frac{1}{2lL}B_p^b} \sin\left[\kappa_{px}\left(x+\frac{L}{2}\right)\right] \\ u_s(x) &= u_s^{Edge}(x) = -\sqrt{\frac{1}{2lL}B_p^e} \sinh\left[\eta\left(x+\frac{L}{2}\right)\right] \\ v_s(x) &= v_s^{Bulk}(x) = \pm\sqrt{\frac{1}{2lL}B_p^b} \sin\left[\kappa_{px}\left(x-\frac{L}{2}\right)\right] \\ v_s(x) &= v_s^{Edge}(x) = \pm i\sqrt{\frac{1}{2lL}B_p^e} \sinh\left[\eta\left(x-\frac{L}{2}\right)\right] \end{aligned} \right\} \quad (14c)$$

where,

$$B_s^b = \left(1 - \frac{\sin(2\kappa_{px}L)}{2\kappa_{px}L}\right)^{-1} \quad (14d)$$

$$B_s^e = e^{\frac{1}{2}\eta L}\left(\frac{\sinh(2\eta L)}{2\eta L} - 1\right)^{-1} \quad (14e)$$

where $l$ represents the normalization length in the y-direction (which goes to infinity in the final result), $A_s$ is a normalization constant, $k_y, k_{xp}$ are the wavenumbers which are defined by the dispersive relation $k_y - ik_{xp} = (k_y + ik_{xp})e^{-2ik_{xp}L}$ and $s = \{p, k_y\}$ is the combinative discrete-continuous index ($p$ is the discrete number with respect to the x-axis and $k_y$ is the continuous wave-number over the *y*-axis). This dispersion relation has an infinite number of real roots with a single point of concentration at infinity (confined modes) and one imaginary root that corresponds to the edge state [16]. Transformation to the edge state from confined modes in Equations (14a)-(14e) and characteristic equation may be done via exchange $\kappa_{px} \to \eta$. The two signs in (14) correspond to electrons and holes respectively. As one can see Eq. (14) satisfies the Dirac equation and is subject to the following boundary conditions; $u_s(-L/2) = v_s(L/2) = 0$ [48]. The components *u(x)* and *v(x)* are separately orthogonal, while non-orthogonal mutually due to their coupling over electron motion between the atoms of A and B sub-lattices. Since the expression for the ac-conductivity with a zigzag edge is isotropic, we have $K_{ab}(\omega;\mathbf{x},\mathbf{x}') = \delta_{ab} K(\omega;\mathbf{x},\mathbf{x}')$, where the general susceptibility is $K(\omega;\mathbf{x},\mathbf{x}') = -i\omega^{-1}(\Pi(\omega) - \Pi(0))$, with

$$\Pi(\omega) = \sum_s \sum_{s'} \frac{(\hat{\mathbf{j}}(\mathbf{x}))_{ss'} (\hat{\mathbf{j}}(\mathbf{x}'))_{s's}}{(\hbar\omega - (\varepsilon_s - \varepsilon_{s'}))} (f(\varepsilon_s) - f(\varepsilon_{s'}))$$ (15)

Here represents the matrix element of the current operator, **σ** denotes the *xy*-vector of the Pauli matrices, $\varepsilon_s = \pm\hbar v_F \sqrt{k_{xp}^2 + k_y^2}$ is the energy of the s-th state and $e, v_F$ correspond to the electron charge and Fermi velocity respectively.

The general susceptibility may be presented here as the sum of two components of different origin $K(\omega;\mathbf{x},\mathbf{x}') = K^{Inter}(\omega;\mathbf{x},\mathbf{x}') + K^{Intra}(\omega;\mathbf{x},\mathbf{x}')$, where $K^{Inter}(\omega;\mathbf{x},\mathbf{x}') = -i\omega^{-1}\Pi(\omega)$ corresponds to the interband motion which may be ignored omitted for rather low (THz) frequencies. The second term $K^{Intra}(\omega;\mathbf{x},\mathbf{x}') = i\omega^{-1}\Pi(0)$ corresponds to the intraband motion and leads to the common conductivity law $\mathbf{j}(\mathbf{x}) = -i\sigma^{Intra}(x)\mathbf{E}(\mathbf{x})$, where the surface conductance is found by substituting the pseudospinor modes (14) into (15), which renders

$$\sigma^{Intra}(x) \approx -\frac{ie^2 v_F^2}{2\pi(\hbar\omega + i0)} \int_{-\infty}^{\infty} \sum_n \frac{\partial f(\varepsilon)}{\partial \varepsilon}\bigg|_{\varepsilon=\varepsilon_s=\varepsilon_n(k_y)} \cdot \left(|u_s(x)|^2 + |v_s(x)|^2\right) dk_y$$ (16)

(for details of deriving Equation (16) see Appendix A). The explicit expression for the conductivity given in (16) is spatially inhomogeneous (*x*-depended) and is a manifestation of edge effect and incorporating a self-consistent description.

*2.3. Approximation for zero temperature.*

Let us next consider the important case of the conductance at zero temperature, which opens the way for further simplification of Equation (16). In this case the Fermi distribution may be replaced by a step function and its derivative in (16) can be transformed into a Dirac delta function $\partial f/\partial\varepsilon = -\delta(\varepsilon - \mu)$, which allows an integration of (16). Performing the integration in possible in terms of the discrete number of roots of the following transcendental equation $\varepsilon(k_y) = \mu$ ( qualitatively depicted in Figure 2), which may be also written as

$$\mathrm{tg}\left(\sqrt{\tilde{\mu}^2 - k_{yn}^2}\, L\right) = \frac{\sqrt{\tilde{\mu}^2 - k_{yn}^2}}{k_{yn}} \quad (17)$$

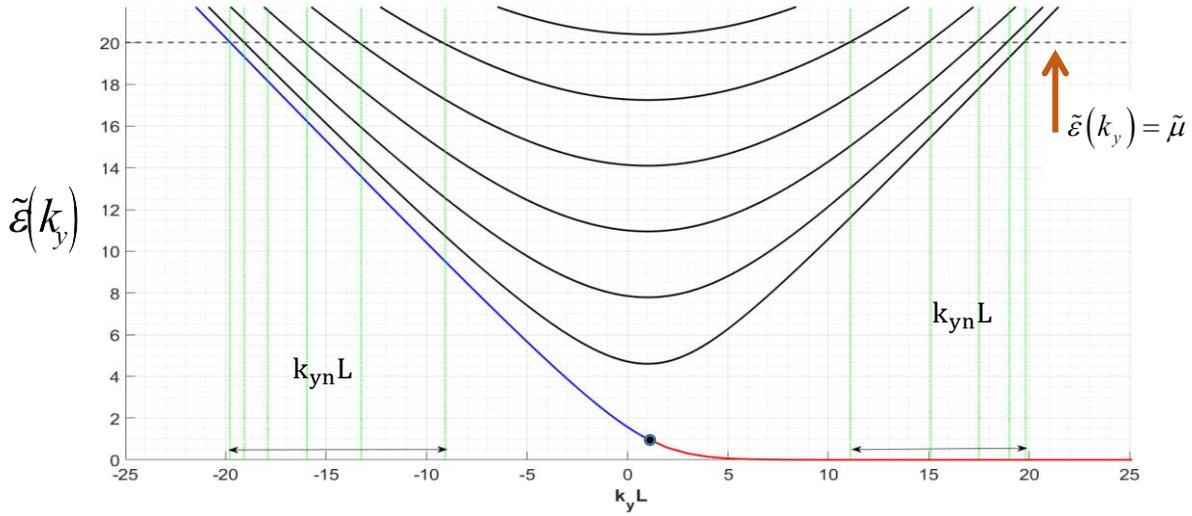

**Figure 2.** Illustration with respect to the zero-temperature approximation. Black lines correspond to confined modes, red line corresponds to the edge mode, and blue line corresponds to the first confined mode. The black point corresponds to the critical wavenumber in which the mutual transformation of edge mode to the first confined mode takes place.

The value $\tilde{\mu} = \mu/\hbar v_F$ means normalizing the electrochemical potential by the electron energy. The corresponding edge mode may be obtained by the formal exchange $k_{yn} \to i\bar{k}$ (such root exists only under special conditions). For the integration over $k_y$ we use the following property of the Dirac delta-function

$$\int_{-\infty}^{\infty} \delta(\varepsilon(k_y) - \mu) F(k_y) dk_y = \sum_n \frac{F(k_{yn})}{|\varepsilon'(k_{yn})|} \quad (18)$$

where $\varepsilon'(k_y)$ denotes the y-component of the group velocity of pseudospin (prime means the derivative). The final explicit result for the conductivity can be written as

$$\sigma^{Intra}(x) = i\frac{e^2 \mu}{\pi \hbar^2 L(\omega + i0)} \sum_{n=1}^{2N+1} \frac{B_n}{|\varepsilon'(k_{yn})|}\left[\sin^2\left(\sqrt{\tilde{\mu}^2 - k_{yn}^2}\left(x - \frac{L}{2}\right)\right) + \sin^2\left(\sqrt{\tilde{\mu}^2 - k_{yn}^2}\left(x + \frac{L}{2}\right)\right)\right] \quad (19)$$

The values of $k_{yn}$ depend on the electrochemical potential and satisfy the characteristic Equation (17). The normalized coefficients $B_n$ in (19) are defined by

$$B_n = \left(1 - \frac{\sin\left(2\sqrt{\tilde{\mu}^2 - k_{yn}^2}L\right)}{2\sqrt{\tilde{\mu}^2 - k_{yn}^2}L}\right)^{-1} = \frac{\tilde{\varepsilon}_n^2}{2\left(\tilde{\varepsilon}_n^2 L - k_{yn}\right)} \tag{20}$$

Equation (19) may be finally transformed to

$$\sigma^{Intra}(x) = i\frac{e^2\mu}{2\pi\hbar^2 L(\omega + i0)} \cdot \sum_{n=1}^{2N+1} \frac{1}{\left(k_{yn}(\tilde{\mu})L - 1\right)}\left[\sin^2\left(\sqrt{\tilde{\mu}^2 - k_{yn}^2}\left(x - \frac{L}{2}\right)\right) + \sin^2\left(\sqrt{\tilde{\mu}^2 - k_{yn}^2}\left(x + \frac{L}{2}\right)\right)\right] \tag{21}$$

(for details see Appendix B).

*2.4. The limit of an infinite sheet.*

In this Subsection we show that our model reduces to the familiar Drude relation for the conductivity in infinitely wide sheet. Taking the limit $L \to \infty$, and using the following transformation $(2L)^{-1}\sum_n(...) \to (2\pi)^{-1}\int_{-\infty}^{\infty}(...)dk_x$. Equation (16) yields

$$\sigma^{Intra}(x) \approx -\frac{ie^2 v_F^2}{2\pi(\hbar\omega + j0)}$$

$$\int_{-\infty}^{\infty}\int_{-\infty}^{\infty}\left.\frac{\partial f(\varepsilon)}{\partial \varepsilon}\right|_{\varepsilon = \varepsilon_s = \varepsilon_n(k_y)} \cdot \left[1 - \frac{1}{2}\cos(k_x(2x - L)) - \frac{1}{2}\cos(k_x(2x + L))\right]dk_x dk_y \tag{22}$$

Introducing the polar variables $k_x = \tilde{\varepsilon}\sin\Phi$, $k_y = \tilde{\varepsilon}\cos\Phi$, we obtain $dk_x dk_y = \tilde{\varepsilon}d\tilde{\varepsilon}d\Phi$. Using the zero temperature approximation, we exchange the Fermi-distribution derivative by the Dirac delta-function and integrate over $\tilde{\varepsilon}$, so that Equation (22) becomes

$$\sigma^{Intra}(x) \approx \frac{ie^2\mu}{2\pi^2\hbar^2(\omega + i0)} \cdot \int_0^{2\pi}\left[1 - \frac{1}{2}\cos(\tilde{\mu}(2x - L)\cos\Phi) - \frac{1}{2}\cos(\tilde{\mu}(2x + L)\cos\Phi)\right]d\Phi \tag{23}$$

Furthermore, using the addtion theorem for Bessel functions [49]

$$e^{\mp i\rho\cos\Phi} = \sum_{m=-\infty}^{\infty} i^{\mp m} J_m(\rho) e^{\pm im\Phi} \tag{24}$$

we integrate (23) and obtain

$$\sigma^{Intra}(x) \approx \frac{ie^2\mu}{\pi\hbar^2(\omega+i0)}\left[1-\frac{1}{2}J_0\left(\tilde{\mu}(2x-L)\right)-\frac{1}{2}J_0\left(\tilde{\mu}(2x+L)\right)\right] \quad (25)$$

If the observation point $x$ (of under the assumption of large L), is placed rather far from the edges, the arguments of Bessel functions become large, so that the asymptotic relations [49]. In this case, the last two terms in (25) become indefinitely small. The first term, which is exactly equal to the Drude conductivity $G^{Drude} = ie^2\mu/\pi\hbar^2(\omega+i0)$, becomes dominate over the whole area of the ribbon excluding the narrow vicinities of the edges. Thus, it is demonstrated that in this limit our model asymptotically reduces to the classical expression for the Drude conductivity.

*2.5. Optical conductance for the edge with infinite-mass boundary condition*

This problem is of special interest in connection with a suspended sheet. The interaction between the edges of the sheet with the electrodes, results in the creation of electrostatic potential. Following [13], the electron-hole symmetry, which is generally restricted for boundary conditions of a zigzag or armchair types is broken. It may be considered as a manifestation of a staggered potential at zigzag boundaries, which may change the nature of the boundary condition. For, infinitely large value of potential it leads to an "infinite-mass" boundary condition, which may be written as $u_s(\pm L/2) = -v_s(\pm L/2)$. The corresponding eigen pseudospins are then given by Equation (14) with

$$u_s(x) = \frac{1}{2\sqrt{2L}}\left(e^{-i\left(k_{xn}x-\frac{\theta_s}{2}\right)} - (-1)^n e^{i\left(k_{xn}x-\frac{\theta_s}{2}\right)}\right) \quad (26)$$

$$v_s(x) = \frac{1}{2\sqrt{2L}}\left(e^{-i\left(k_{xn}x+\frac{\theta_s}{2}\right)} - (-1)^n e^{i\left(k_{xn}x+\frac{\theta_s}{2}\right)}\right) \quad (27)$$

where $k_{xn} = n\pi/L$ and $e^{i\theta_s} = \frac{k_y + ik_{xn}}{\sqrt{k_y^2 + k_{xn}^2}}$.

In order to calculate the conductance, we use the Kubo approach with the pseudo spinors defined in (26) and (27). The difference from the zigzag edge formulation, is due to the lack of orthogonality inf (26) and (27), which leads to a non-local conductance (spatial dispersion). The conductance operator does not add up to the convolution form because of the non-homogeneity of the finite-length structure. In the case of a rather wide sheet the non-local component becomes relatively small and may be usually ignored. In this particular case the conductance relates to Equation (16) with the pseudospins given given in (26) and (27).

**3. Numerical results and discussion**

The optical properties of graphene are generally defined by the geometrical size of the sample and the value of the electrochemical potential. These parameters may vary over a wide range and thus are of practical interest to nanoantenna design. The recent advances in graphene technology make it possible to produce graphene samples from few nanometers to few centimeters [6,7,21]. The electrochemical potential may also vary over the interval $0 < \mu < 1.0$ eV, by doping the sample or by applying gate voltage [6,7,21]. In this Section we will present some numerical results of conductivity simulations for a wide range of physical parameters, based on the theory developed above.

One of the main results of the present study according to Equations (16) and (19), is that the non-homogeneity of conductance is mainly due to the edge effects. The conductance distribution is controlled by the parameter $\tilde{\mu}L = \mu L/\hbar v_F$, which defines the number of modes supported by the conductance value. Figures 3-5 present the normalized conductance distribution for a rather large length $L = 800$ nm and for different values of the electrochemical potential (increasing $\mu$ corresponds to increasing number of modes). The qualitative behavior of the conductance is the same in all these Figures. The conductance oscillates with respect to the spatial variable and decreases near the edges. The amplitude and period of oscillations decrease with increasing of the electrochemical potential $\mu$. The average value of the conductance (to a high degree of accuracy), corresponds to the classical Drude model of conductivity, as discussed in Section 2.4. One can anticipate that such small oscillations are unable to manifest themselves in the scattering of electromagnetic waves, due to the smallness of their period compared with the wavelength. Thus, we may conclude that the Drude model can be applied for this range of parameters.

The considered scenario changes dramatically with shortening of the sheet, as shown in Figures 6-8 (each Figure corresponds to a different value of the length for the same value of electrochemical potential). One can clearly see the enhancement of oscillations, which makes the Drude model invalid for such values of parameters. In fact, it becomes impossible to introduce the conductivity concept in its ordinary meaning. As it was mentioned above, the value defined by Equation (16) has the meaning of optical conductivity, which strongly depend on the geometrical size of the sheet. It may be coupled with Drude conductivity by the Relation

$$\sigma^{Intra}(x) = G^{Drude} \frac{1}{L} \Theta(x) \tag{28}$$

where

$$\Theta(x) = \sum_{n=1}^{2N+1} \frac{1}{(k_{yn}(\tilde{\mu})L - 1)} \left[ \sin^2\left(\sqrt{\tilde{\mu}^2 - k_{yn}^2}\left(x - \frac{L}{2}\right)\right) + \sin^2\left(\sqrt{\tilde{\mu}^2 - k_{yn}^2}\left(x + \frac{L}{2}\right)\right) \right] \tag{29}$$

is $x$-dependent coefficient, in which the configuration of the sample manifests itself. The situation is rather similar to the electron transport in graphene in dc field (the concepts of conductance and conductivity discussed and compared in [16,50]).

The physical mechanism for the conductivity oscillations is the interference between the pseudospin modes due to the reflection from the sheet boundaries. The important features are demonstrated in the single-mode conductance (Figures 6 (a),(b)). For the rather small electrochemical potential the active mode is an edge type. The sign of the conductance is negative, which corresponds to its inductive origin. The increase in the electrochemical potential leads to the transformation from inductive to capacitive one (sign exchange) due to the transformation from the edge mode to the bulk one.

As one can see, the conductivity of a graphene sheet changes its qualitative behavior for rather small values of length. However, graphene antennas generally exhibit a resonate behavior at much lower frequencies as well as their metallic counterparts, which is experimentally implemented in the THz range [43,45,46,51]. Thus one can effectively exploit the electrically tunable conductance of graphene exactly for such small sizes, where the conventional models of conductivity become invalid due to the importance of edge effects. In summary, it is important to note that including edge effects in the physical modeling, opens a new way for electrical controlling of resonant graphene antennas via the overturned electrochemical potential by means of the gate voltage. In some cases where the electrochemical potential varies adiabatically slow in time, it will also produce a modulation of the THz emission.

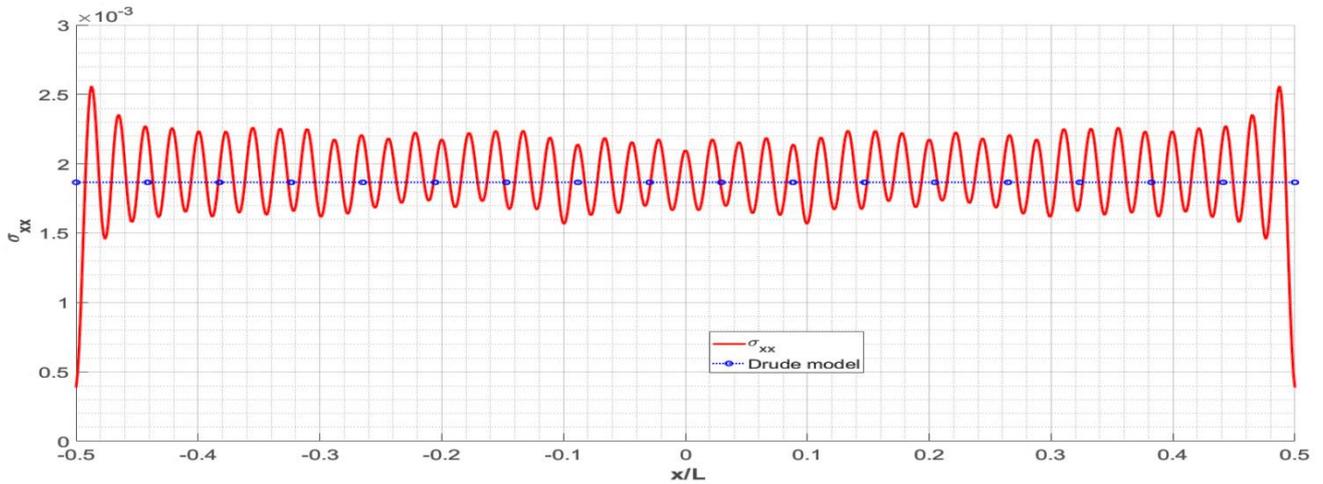

**Figure 3.** The spatial distribution of the conductance in the units $G_{Drude}/Ll$. $L$=800nm, $\mu$=0.1eV.

The concept of the optical conductance developed in this paper allows formulating the effective boundary conditions for electromagnetic field at the surface of graphene sheet in the form

$$\left[\mathbf{n},[\mathbf{H}_{z=+0}-\mathbf{H}_{z=-0},\mathbf{n}]\right] = G^{Drude}\Theta(x)[\mathbf{n},\mathbf{E}] \qquad (30)$$

Their using leads to modification of integral equations of antenna theory and methods of their solution.

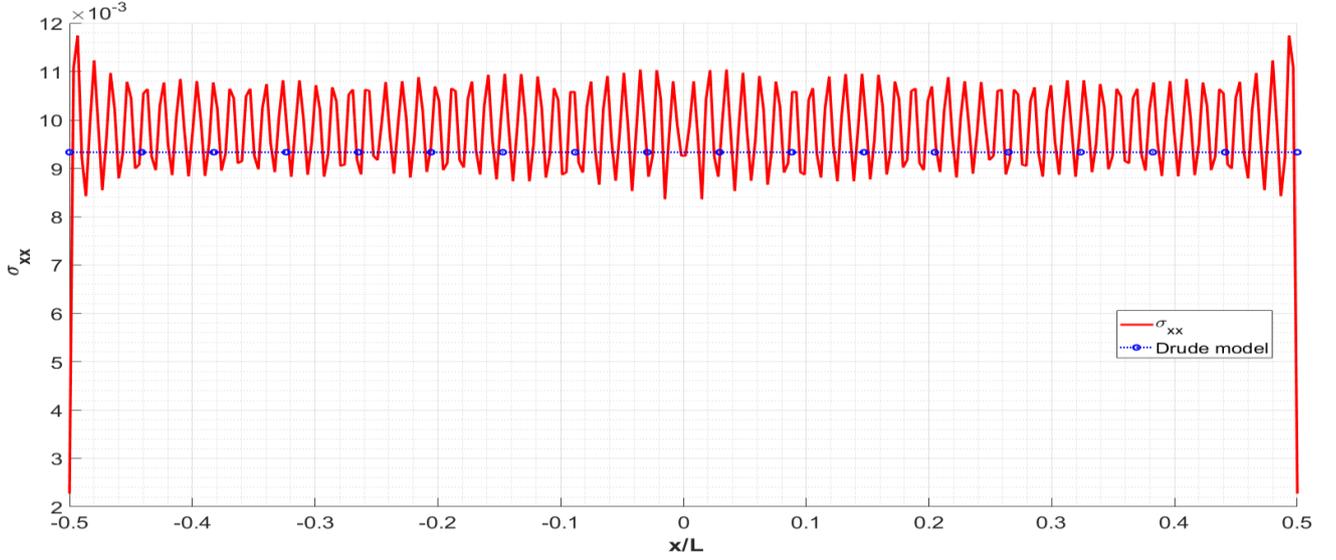

**Figure 4.** The spatial distribution of the conductance in the units $G_{Drude}/Ll$. $L$=800nm, $\mu$=0.5eV.

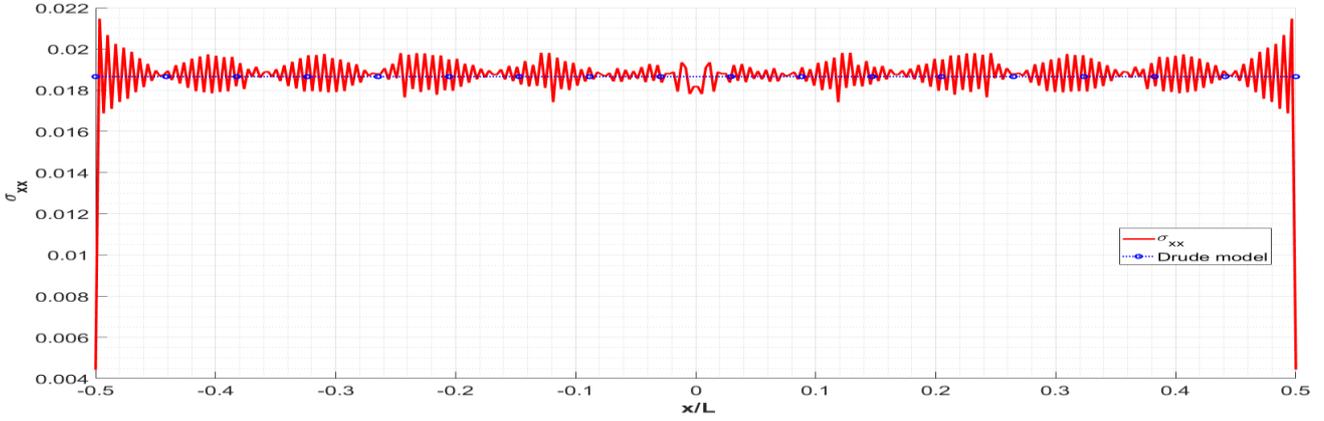

**Figure 5.** The spatial distribution of the conductance in the units $G_{Drude}/Ll$. L=800nm, $\mu$=1.0eV.

The considered scenario changes dramatically with shortening of the sheet, as shown in Figures 6-8 (each Figure corresponds to a different value of the length for the same value of electrochemical potential). One can clearly see the enhancement of oscillations, which makes the Drude model invalid for such values of parameters. In fact, it becomes impossible to introduce the conductivity concept in its ordinary meaning. The value defined by Equation (16), has the meaning of an "effective" conductivity, which strongly depend on the geometrical size of the sheet. The situation is rather similar to attempt to describe the optical properties of semiconductor quantum dots via the dielectric function in the limit of weak conferment [50] (the "effective" dielectric function strongly depends on the sample configuration). The physical mechanism for the conductivity oscillations, is the interference between the pseudospin modes due to the reflection from the sheet boundaries. The important features are demonstrated in the single- mode conductivity (Figures 6 (a), (b)). For the rather small electrochemical potential the active mode is an edge type. The sign of the conductivity is negative, which corresponds to its inductive origin. The increase in the electrochemical potential leads to the transformation from inductive to capacitive one (sign exchange), due to the transformation from the edge mode to the bulk one.

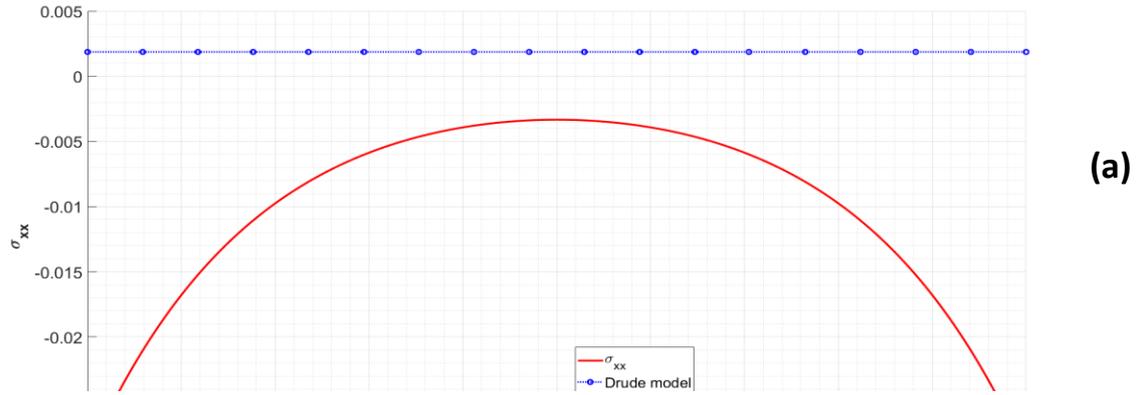
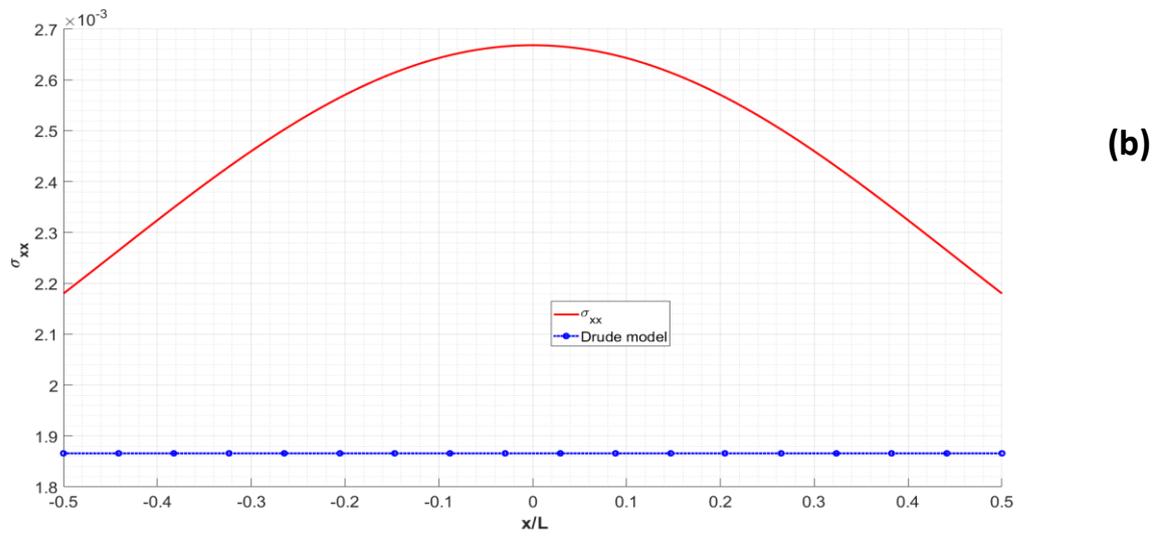
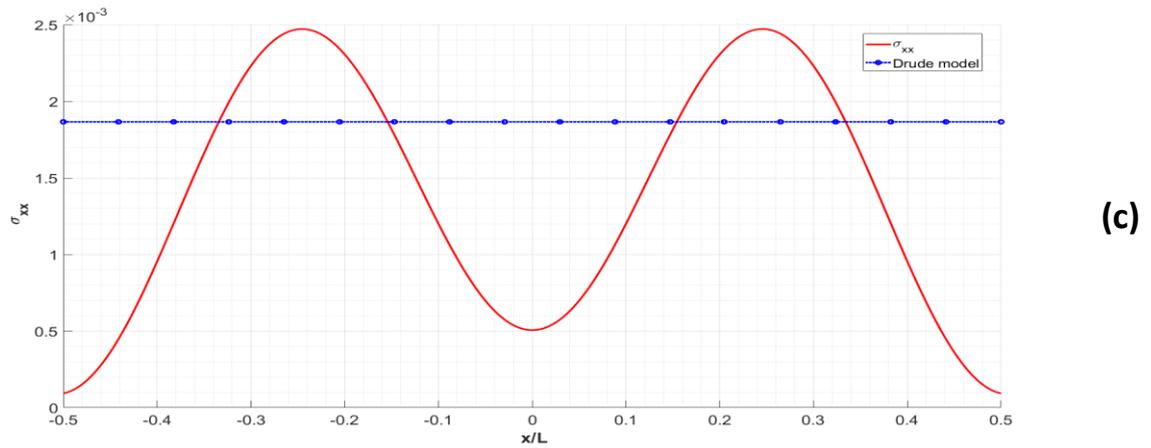

**Figure 6.** The spatial distribution of the conductance in the units $G_{Drude}/Ll$ for different values of length and $\mu=0.1$ eV; **(a)** L=2nm (single-mode regime; edge mode); **(b)** L=10nm (single-mode regime; bulk mode); **(c)** L=50nm (three-mode regime; all modes are of bulk type).

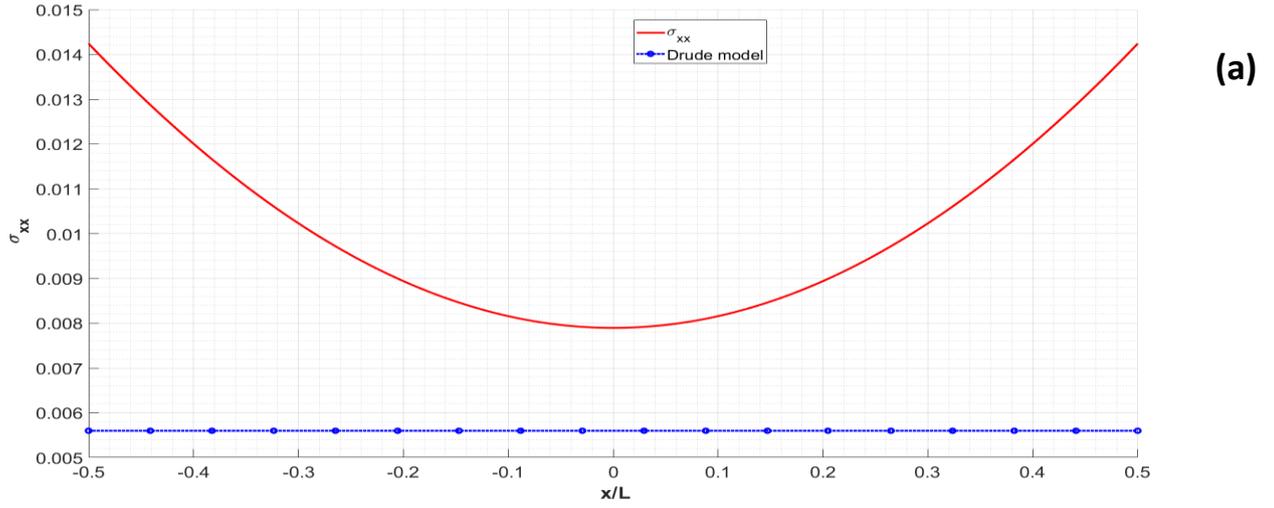

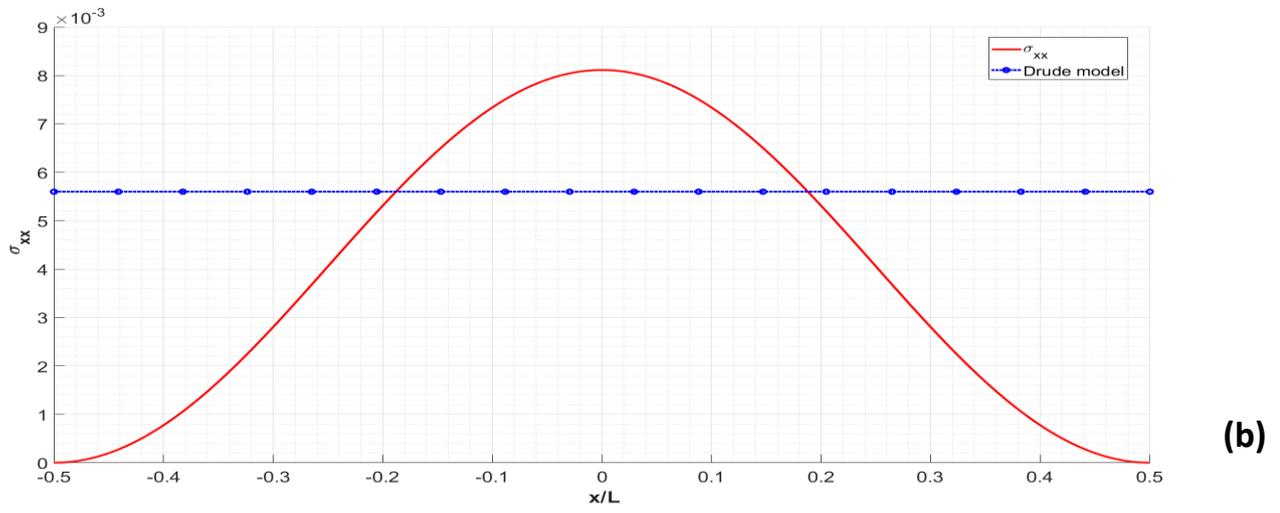

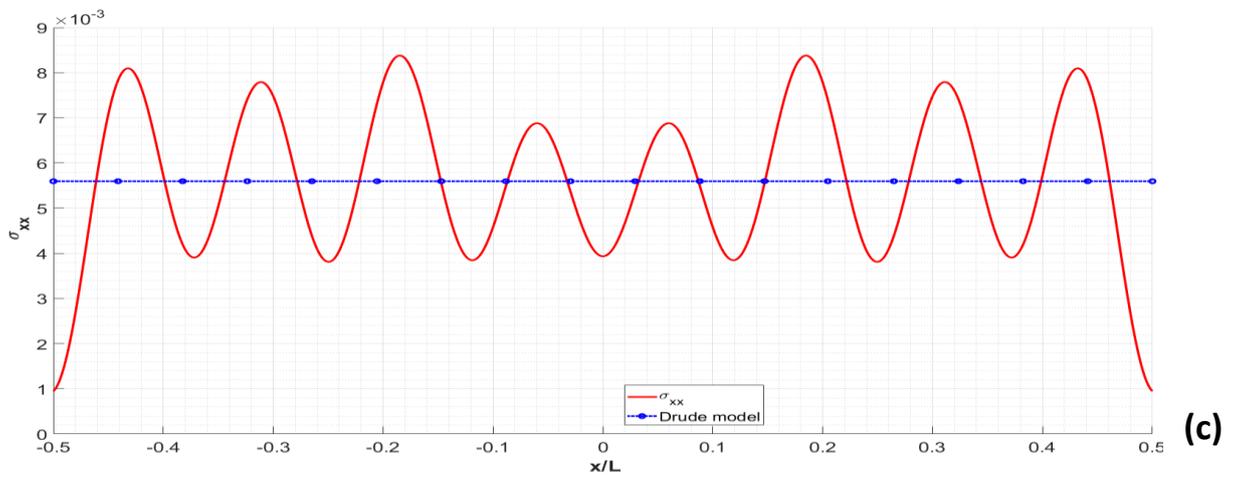

**Figure 7.** The spatial distribution of the conductance in the units $G_{Drude}/Ll$ for different values of length. $\mu$ =0.3 eV; **(a)** L=2nm; **(b)** L=10nm; **(c)** L=50nm.

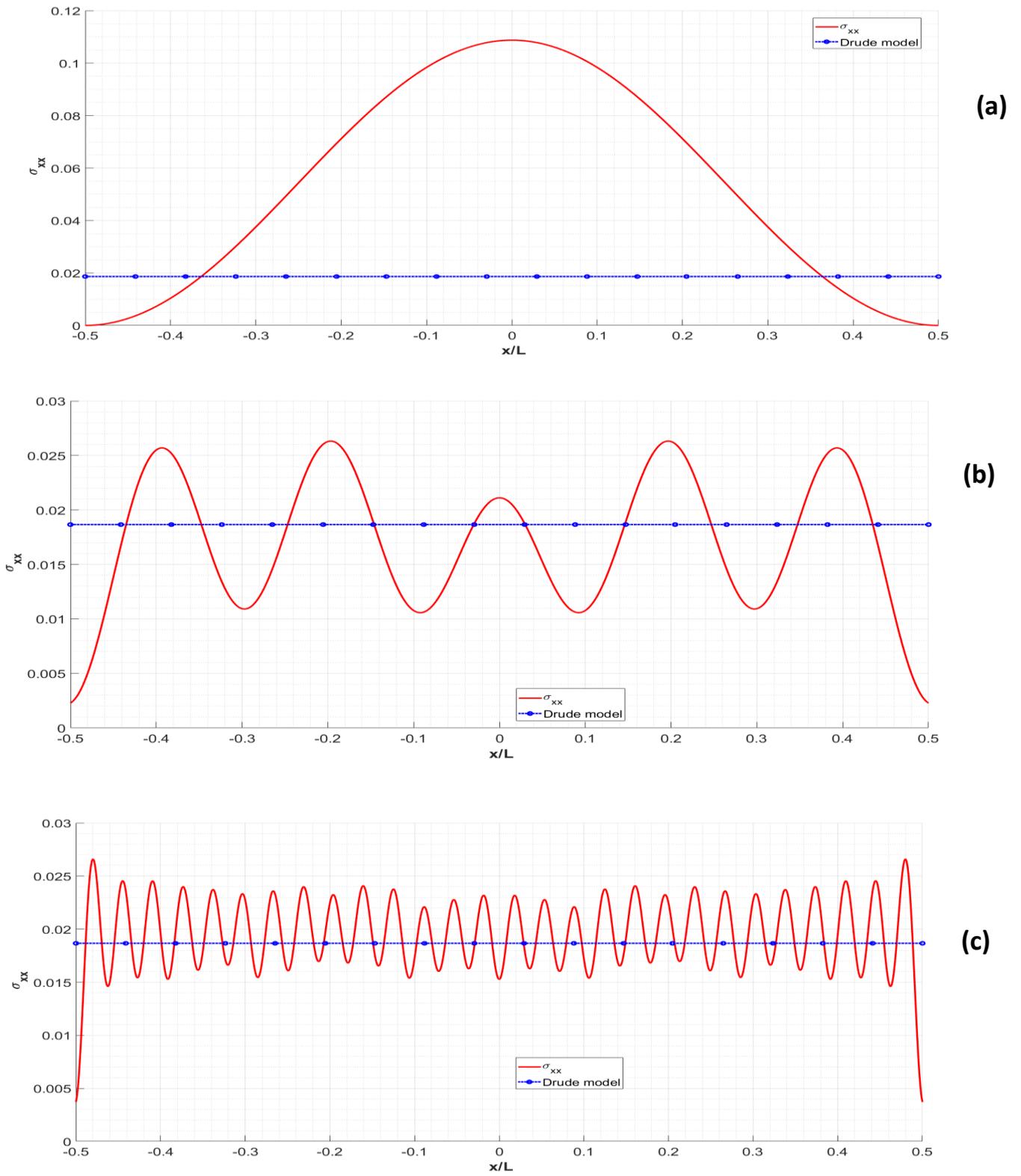

**Figure 8.** The spatial distribution of the conductance in the units $G_{Drude}/Ll$ for different values of length. $\mu = 1.0$ eV; **(a)** L=2nm; **(b)** L=10nm; **(c)** L=50nm.

# 4. Conclusion and outlook.

The main results of the paper can be summarized as follows:

1) We have developed a new theory of interaction of electromagnetic field with graphene sheet for nanoantenna applications in the THz, infrared and optical frequency ranges. The main characteristic feature of our theory is accounting for edge effects in a self-consistent manner. It is based on the concept of optical conductance considered as a general susceptibility and calculated by Kubo approach. The model is based on the concept of Dirac pseudo-spins founded via solving the boundary-value problem for the Dirac equation with the appropriate boundary conditions satisfying the physical model, including edge effects of the sheet;
2) The main manifestation of the importance of edge effects is demonstrated by the inhomogeneity of the optical conductance. The amplitude and period of its oscillations depend on the length of the sheet and on the electrochemical potential. It is defined by the number of pseudo-spin modes supporting the conductance;
3) The developed theory is applied for the simulation of the sheet conductance in a wide range of sample parameters ( length 2.1nm – 800nm and electrochemical potential 0.1 – 1.0 eV). It is shown, that for a length exceeding 800nm our model and the widely used Drude model of conductivity agree to a high degree of accuracy. However, for small geometric sizes (i.e., smaller than 50nm), the physical picture of conductivity with respect to the Drude model changes dramatically due to the influence of edge effects. This circumstance should be accounted for in the design of graphene-based resonant THz antennas and other types of photonic and plasmonic nanodevices;
4) It is shown, that the qualitative distribution of the conductivity along the sheet strongly depends on the electrochemical potential. Thus, it is possible to control the conductivity and performance of graphene nanoantennas, by means of varying the gate voltage;

Our theory allows reformulation of the effective boundary conditions for the electromagnetic field at the surface of the graphene sheet with accounting of the edge effects. It requires the modification of integral equations of antenna theory and the methods of their solution. This should be one of the subjects of future research activity as well as their application to nanoantennas and other nanodevices.

**Author Contributions:** Developments of the physical models, derivation of the basis equations, interpretation of the physical results and righting the paper have been done by T.B., T.M., O.G. and G.S. jointly. The numerical simulations were produced by T.B.

**Funding:** This research was funded by NATO grant number NATO SPS-G5860 and by H2020,project TERASSE 823878.

**Institutional Review Board Statement:** Not applicable.

**Informed Consent Statement:** Not applicable.

**Data Availability Statement:** Not applicable.

**Conflicts of Interest:** The authors declare no conflict of interest

**Appendix A. Derivation of Equation (16).**

In this Appendix we discuss the boundary-value problem for pseudo-spin defined by Equation (14). As it was mentioned above, the pseudo-spin satisfies the Dirac equation with the following boundary conditions; $u_s(-L/2) = v_s(L/2) = 0$ [25]. It may be also transformed into the Helmholtz equation with two special sets of boundary conditions. We have the Dirichlet condition at the left-hand side and the impedance condition $(k_y - \partial_x)u|_{x=L/2} = 0$ at the right-hand side for the components $u(x)$ (determined by the Dirac Equation). The situation is precisely inverted for the second type, namely a Dirichlet condition at the right-hand side and an impedance condition $(k_y + \partial_x)v|_{x=-L/2} = 0$ at the left-hand side. These problems are Hermitian, whereby the eigenmodes form a complete basis. The components $u(x)$ and $v(x)$ are both separately orthogonal, but are mutually non-orthogonal, due to their coupling over the electron motion between the atoms of A and B sublattices. The property of orthogonality is shown at Appendix C. Using completeness, orthogonality and normalization conditions $\int_{-L/2}^{L/2} |u_p(x)|^2 dx + \int_{-L/2}^{L/2} |v_p(x)|^2 dx = 1$, we obtain

$$2\sum_{p'} u_{p'}^*(x) u_{p'}(x') = \delta(x-x') \tag{A.1}$$

$$2\sum_{p'} v_{p'}^*(x) v_{p'}(x') = \delta(x-x') \tag{A.2}$$

Starting from the *xx*-component and using the basis relation (11) for *a=x, b=x*, the matrix element of the current density operator, one gets

$$\left(\hat{j}_x(\mathbf{x},0)\right)_{ss'} = \frac{1}{l} e v_F \left(u_{n'}(x) v_n(x) + v_n(x) u_{n'}(x)\right) e^{i(k_y - k_{y'})y} \tag{A.3}$$

Next we examine the limit of $l \to \infty$ by making the exchange $l^{-1}\sum_s \to (2\pi)^{-1}\int_{-\infty}^{\infty} \sum_n$ and the same for $s', n'$. Summing over $s,s'$ and taking into account both electrons and holes ($\pm i n\, v_n(x)$) as well and the charge carriers in two valleys K and K', leads to.

$$K_{xx}(\mathbf{x},\mathbf{x}';\omega) \approx -\frac{ie^2 v_F^2}{4\pi^2(\omega+i0)} \int_{-\infty}^{\infty}\int_{-\infty}^{\infty} dk_y dk_y' \cdot e^{i(k_y - k_y')(y-y')} \cdot$$
$$\sum_n \left.\frac{\partial f(\varepsilon)}{\partial \varepsilon}\right|_{\varepsilon=\varepsilon_n(k_y)} \cdot \sum_{n'} \Lambda_{nn'}^{(xx)}(\mathbf{x},\mathbf{x}';k_y) \tag{A.4}$$

where

$$\sum_{p'}\Lambda^{(xx)}_{pp'}(x,x';k_y)=$$

$$v_p(x)v_p(x')\sum_{p'}u_{p'}(x)u_{p'}(x')+$$

$$u_p(x)u_p(x')\sum_{p'}v_{p'}(x)v_{p'}(x')+ \qquad (A.5)$$

$$u_p(x)v_p(x')\sum_{p'}v_{p'}(x)u_{p'}(x')+$$

$$v_p(x)u_p(x')\sum_{p'}u_{p'}(x)v_{p'}(x')$$

Note, that the summation in (A.5), means including the contribution from both electrons and holes and the valleys K, K'. The sum over electrons and holes may be transformed to the sum over the electron states by means of their doubling. The sum of the two other terms is zero due to the opposite sign of $v_n(x)$ (subject to the same $u_n(x)$). The sums over the electron states are decomposed into the components over the two valleys, implying that

$$\sum_n(...)=\sum_n{}_K(...)+\sum_n{}_{K'}(...)=2\sum_n{}_K(...) \qquad (A.6)$$

Finally invoking these transformations and using the well-known identity $(2\pi)^{-1}\int_{-\infty}^{\infty}e^{ihy}dh=\delta(y)$, we obtain

$$K_{xx}(\omega;\mathbf{x},\mathbf{x}')\approx -\frac{ie^2 v_F^2}{2\pi(\omega+i0)}\delta(\mathbf{x}-\mathbf{x}')\int_{-\infty}^{\infty}dk_y\sum_n\left.\frac{\partial f_0(\varepsilon)}{\partial \varepsilon}\right|_{\varepsilon=\varepsilon_n(k_y)}\cdot\left(u_n^2(x)+v_n^2(x)\right)$$

(A.7)

The above equation relates to the only existing spin-state (a real physical spin rather than a pseudospin). Therefore, the total conductivity must be doubled, which corresponds to the value of the conductivity given by Equation (16). Other components of the conductivity tensor may be obtained in a similar way. For example, we have $K_{yy}(\omega;\mathbf{x},\mathbf{x}')=K_{xx}(\omega;\mathbf{x},\mathbf{x}')$ and $K_{xy}(\omega;\mathbf{x},\mathbf{x}')=K_{yx}(\omega;\mathbf{x},\mathbf{x}')=0$.

**Appendix B: Group velocity of pseudospins in graphene sheet.**

Here we calculate the normalized group velocity of the pseudospins defined as $v_g=\tilde{\varepsilon}'(k_y)=\partial\tilde{\varepsilon}/\partial k_y$, based on the characteristic equation $f(k_x)=\frac{\text{tg}(k_x L)}{k_x}=k_y^{-1}$. The y-component of the wavevector is considered as an independent variable. By taking derivative with respect to $k_y$, one gets

$$\frac{\partial f}{\partial k_y} = \frac{\partial f}{\partial k_x} \cdot \frac{\partial k_x}{\partial k_y} = -\frac{1}{k_y^2} \tag{A.8}$$

And by using the relation $k_x = \sqrt{\tilde{\varepsilon}^2(k_y) - k_y^2}$, we obtain

$$\frac{\partial k_x}{\partial k_y} = \frac{\tilde{\varepsilon}\frac{\partial \tilde{\varepsilon}}{\partial k_y} - k_y}{\sqrt{\tilde{\varepsilon}^2(k_y) - k_y^2}} = \frac{\tilde{\varepsilon} v_g - k_y}{\sqrt{\tilde{\varepsilon}^2(k_y) - k_y^2}} \tag{A.9}$$

For the derivative $\partial f / \partial k_x$ we have

$$\frac{\partial f}{\partial k_x} = \frac{\frac{k_x L}{\cos^2(k_x L)} - \mathrm{tg}(k_x L)}{k_x^2} \tag{A.10}$$

The trigonometric functions may be expressed through the algebraic ones using the characteristic equation which gives

$$\frac{\partial f}{\partial k_x} = \frac{\tilde{\varepsilon}^2(k_y) L - k_y}{k_y^2 k_x} \tag{A.11}$$

Expressing the group velocity from (A.9) and using (A.10), (A.11), we obtain

$$v_g(k_y) = \frac{\tilde{\varepsilon}(k_y)(k_y L - 1)}{(\tilde{\varepsilon}^2(k_y) L - k_y)} \tag{A.12}$$

In order to determine the renormalization coefficient $B_n$, we can make a similar transformation: express $\sin(2k_x L)$ thought $\mathrm{tg}(k_x L)$ and apply the characteristic Equation (17), which renders by virtue of Equation (20),

$$B_n \left( \frac{\partial \tilde{\varepsilon}}{\partial k_y} \bigg|_{k_y = k_{yn}} \right)^{-1} = \frac{\tilde{\varepsilon}(k_{yn})}{2(k_{yn} L - 1)} \tag{A.13}$$

with the conductivity given explicitly in Equation (21).

**Appendix C: Orthogonality of pseudo-spin modes**

The Equations for pseudo-spin mode with number $s$ has the form

$$\left. \begin{array}{l} \dfrac{\partial v_s}{\partial x} = -k_y v_s - \tilde{\varepsilon}_s u_s, \\ \dfrac{\partial u_s}{\partial x} = k_y u_s + \tilde{\varepsilon}_s v_s \end{array} \right\}, \tag{A.14}$$

The similar Relation may be formulated for another mode with the number $s'$:

$$\left.\begin{aligned}\frac{\partial v_{s'}}{\partial x} &= -k_y v_{s'} - \tilde{\varepsilon}_{s'} u_{s'}, \\ \frac{\partial u_{s'}}{\partial x} &= k_y u_{s'} + \tilde{\varepsilon}_{s'} v_{s'}\end{aligned}\right\} \tag{A.15}$$

Let us multiply first Equation (A.14) to $u_{s'}$ and second Equation multiply to $v_{s'}$. Summarize them and integrate over the interval $-L/2 < x < L/2$. Using the boundary conditions, we obtain

$$\tilde{\varepsilon}_s \int_{-L/2}^{L/2} u_s(x) u_{s'}(x) dx - \tilde{\varepsilon}_{s'} \int_{-L/2}^{L/2} v_s(x) v_{s'}(x) dx = 0 \tag{A.16}$$

The similar Relation may be obtained from (A.16) by rearranging the indexes $s$ and $s'$. We have

$$\tilde{\varepsilon}_{s'} \int_{-L/2}^{L/2} u_s(x) u_{s'}(x) dx - \tilde{\varepsilon}_s \int_{-L/2}^{L/2} v_s(x) v_{s'}(x) dx = 0 \tag{A.17}$$

The eigenvalues with the different indexes are non-degenerate. The pair of Equations (A.16),(A.17) may be considered as a system of linear algebraic equations with respect to the integrals. The determinant of this system $\begin{vmatrix}\tilde{\varepsilon}_s & -\tilde{\varepsilon}_{s'} \\ \tilde{\varepsilon}_{s'} & -\tilde{\varepsilon}_s\end{vmatrix}$ is non-zero. It means, that for $s \neq s'$

$$\int_{-L/2}^{L/2} u_s(x) u_{s'}(x) dx = \int_{-L/2}^{L/2} v_s(x) v_{s'}(x) dx = 0 \tag{A.18}$$

which gives the orthogonality relation in the required form.

# References

1. Novotny, L.; van Hulst, N. F. Antennas for light, *Nat. Photonics*, **2011**, 5,83 – 90.
2. Giannini, V.; Fernandez-Domínguez, A.I.; Heck, S.C.; Maier,S.A. Plasmonic nanoantennas: fundamentals and their use in controlling the radiative properties of nanoemitters, *Chem.Rev.***2011**, 111,3888–3912.
3. Bharadwaj, P.; Deutsch, B.; Novotny, L. Optical antennas, *Advances in Optics and Photonics* ,**2009**, 438–483.
4. Parzefall, M.; Novotny, L. Optical antennas driven by quantum tunneling: a key issues review, *Rep. Prog. Phys.*, **2019**,82 ,112401.
5. Slepyan, G.Y.; Vlasenko, S.; Mogilevtsev, D. Quantum Antennas. *Adv. Quantum Technol.***2020**, 3, 1900120. [Cross Ref]
6. Ullah, Z.; Witjaksono, G.; Nawi, I.; Tansu, N.; Khattak, M. I.; Junaid, M. A review on the development of tunable graphene nanoantennas for terahertz optoelectronic and plasmonic applications, *Sensors*, **2020** ,20,1401.
7. Baydin, A.; Tay, F.; Fan, J.; Manjappa, M.; Gao, W.; Kono, J. Carbon nanotube devices for quantum technology, *Materials*,**2022,**15, 1535.
8. Pistolesi, F.; Cleland, A. N.; Bachtold, A. Proposal for a nanomechanical qubit, *Phys. Rev. X* ,**2021**, 11,031027.
9. De Volder, M.F.L.; Tawfick,S.H.; Baughman, R.H.;Hart,J.H. Carbon nanotubes: present and future commercial applications, *Science*, **2013**,V 3391.
10. Basov, D.N.; Fogler, M.M.; Lanzara A.; Feng Wang; Yuanbo Zhang. Colloquium:Graphene spectroscopy, *Rev. Mod. Phys.*,**2014**,86,959.

11. Kavitha, S.; Sairam, K. V. S. S. S. S.; Singh, A. Graphene plasmonic nano-antenna for terahertz communication, *Springer Nature, Applied Sciences*, **2022**, 4114.
12. Hofferber, E.M.; Stapleton, J.A.; Iverson, N.M. Single walled carbon nanotubes as optical sensors for biological applications, *Journal of The Electrochemical Society,* **2020**, 167,037530.
13. Nadeem, S.; Ijaz, S. Single wall carbon nanotube (SWCNT) examination on blood flow through a multiple stenosed artery with variable nanofluid viscosity, *AIP Advances*, **2015**, 107217,5.
14. Margetis, D.; Stauber, T. Theory of plasmonic edge states in chiral bilayer systems, *Phys. Rev. B*, **2021**, 104,115422.
15. Yao, W.; Shengyuan, A. Y.; Niu,Q. Edge states in graphene: from gapped flat-band to gapless chiral modes, *Phys. Rev. Lett.*, **2009**, 102,096801.
16. Katsnelson, M. I. *The physics of graphene,* 2nd ed. Cambridge University Press, Cambridge, 2020.
17. Felsen, L. B.; Marcuvitz, N.; *Radiation and scattering of waves*, IEEE Press series on electromagnetic waves, originally published, Prentice-Hall, Englewood Cliffs, N.J.,1972.
18. Noble, B, *The Wiener-Hopf technique*. 2nd edition, Chelsea Pub. Co. 1988.
19. Miloh,T. Opto-electro-fluidics and tip coax conical surface plasmons, *Phys. Rev. Fluids*, ,**2016**, 1, 044105 .
20. Meixner, J. The behavior of electromagnetic fields at edges, *IEEE Trans. Antennas Propag.*, **1972**, 20, 442-446.
21. Castro Neto, A.H.; Guinea, F.; Peres, N. M. R.; Novoselov, K. S.; Geim, A. K. The electronic properties of graphene, *Rev. Mod. Phys*. **2009**, 81, 109-135.
22. Armitage, N. P.; Mele, E. J.; Vishwanath,V. Weyl and Dirac semimetals in three-dimensional solids, *Rev. Mod. Phys.* **2018**, 90, 015001.
23. Qi, X.L,; Zhang,S.C. Topological insulators and superconductors, *Rev. Mod. Phys.,* **2011**, 83,1057.
24. Hasan, M. Z.; Kane, C. L. Topological insulators, *Rev. Mod. Phys.* **2010,** 82, 3045.
25. Akhmerov, A. R.; Beenakker, C. W. J. Boundary conditions for Dirac fermions on a terminated honeycomb lattice, *Phys. Rev. B*, **2008**, 77, 085423.
26. Balanis,C.A. *Antenna Theory-Analysis and Design,* Third Edition, A John Wiley & Sons, Inc., Publication, 2005.
27. Agio, M.; Alu, A. *Optical Antennas*, Cambridge University Press, Cambridge, 2013.
28. Slepyan, G.Y.; Shuba, M.V.; Maksimenko, S.A.; Lakhtakia, A. Theory of optical scattering by achiral carbon nanotubes and their potential as optical nanoantennas, *Phys. Rev. B*. **2006**, 73,195416.
29. Shuba, M.V.; Slepyan, G.Y.; Maksimenko, S.A.; Thomsen, C.; Lakhtakia, A. Theory of multiwall carbon nanotubes as waveguides and antennas in the infrared and the visible regimes, *Phys. Rev. B*. **2009**, 79,155403.
30. Hanson, G.W.; Radiation efficiency of nano-radius dipole antennas in the microwave and far-infrared regimes, *IEEE Antennas Propag. Mag.* **2008**,50, 66–77.
31. Hao, J.; Hanson, G.W. Infrared and Optical Properties of CarbonNanotube Dipole Antennas, *IEEE Transactions on Nanotechnology* **2006**, 5, 6, 766.
32. Slepyan, G.Y.; Maksimenko, S.A.; Lakhtakia, A.; Yevtushenko,V; Gusakov,V. Electrodynamics of carbon nanotubes: Dynamics conductivity, impedance boundary conditions and surface wave propagation *Phys. Rev. B*, **1999**, 60, 17136–17149.
33. Shuba, M.V.; Paddubskaya, A.G.; Plyushch, A.O.; Kuzhir, P.P.; Slepyan, G.Y.; Maksimenko, S.A.; Ksenevich, V.K.; Buka, P.; Seliuta, D.; Kasalynas, I.; Macutkevic, J.; Valusis, G. L; Thomsen, C.; Lakhtakia, A. Experimental evidence of localized plasmon resonance in composite materials containing single-wall carbon nanotubes, *Phys. Rev. B*. **2012**, 85 ,165435.
34. Slepyan, G.Y.; Shuba, M.V.; Maksimenko, S.A.; Thomsen, C.; Lakhtakia, A. Terahertz conductivity peak in composite materials containing carbon nanotubes: theory and interpretation of experiment, *Phys. Rev. B*.**2010,** 81,205423.

35. Hanson, G.W. Fundamental transmitting properties of carbon nanotube antennas, *IEEE Trans. Antennas Propag.* **2005**,53,3426–3435.
36. Hanson, G.W. Dyadic Green's functions and guided surface waves for a surface conductivity model of graphene, *Journal of Applied Physics*, **2008**, 103, 064302.
37. Falkovsky, L.A. Optical properties of graphene, *Journal of Physics: Conference Series*, **2008**,129, 012004.
38. Falkovsky, L.A.; Varlamov, A.A. Space-time dispersion of graphene conductivity, *Eur. Phys. J. B*, **2007**, 56, 281.
39. Falkovsky, L.A.; Pershoguba, S.S. Optical Far-Infrared Properties of a Graphene Monolayer and Multilayer, *Phys. Rev.B*,2007, 76, 153410
40. Falkovsky, L.A. Optical properties of graphene and IV±VI semiconductors, *Uspekhi Fizicheskikh Nauk*,**2008**, 178-9, 923-934.
41. Vorobev, A.S.; Bianco,G.V.; Bruno,G.; D'Orazio,A.; O'Faolain,L.;Grande,M. Tuning of Graphene-Based Optical Devices Operating in the Near-Infrared, *Appl. Sci.*,**2021**,11, 8367.
42. Gosciniak,J.; Tan, D.T.H. Graphene-based waveguide integrated dielectric-loaded plasmonic electro-absorption modulators, *Nanotechnology,***2013**, 24,185202, 9.
43. Kosik,M.; Müller,M.M.; Słowik,K.; Bryant,G.; Ayuela,A.;Rockstuhl,C; Pelc,M. Revising quantum optical phenomena in adatoms coupled to graphene nanoantennas, *Nanophotonics,* **2022,** 11(14), 3281–3298.
44. Calafel,A.; Cox, J.D.;Radonjić,M.; Saavedra, J. R. M. ; García de Abajo, F. J.; Rozema, L. A. ; Walther, P. Quantum computing with graphene plasmons, *Quantum Information*, **2019,** 37**.**
45. Ye, M.; Crozier, K.B. Metasurface with Metallic Nanoantennas and Graphene Nanoslits for Sensing of Protein Monolayers and Sub-Monolayers, *Optics Express,* **2020**,18479, 28, 12.
46. Singh,A.; Andrello, A.; Thawdar,N.; Jornet, J.M. Design and Operation of a Graphene-Based Plasmonic Nano-Antenna Array for Communication in the Terahertz Band, *IEEE Journal on Selected Areas in Communications*, **2020**, 2104, 38, 9.
47. Landau, L. D.; Lifshitz, E. M.; *Statistical Physics (Course of Theoretical Physics), Pergamon*, New York, Vol. 5, 1980.
48. Brey, L.; Fertig, H.A. Electronic States of Graphene Nanoribbons Studied with the Dirac Equation, *Phys. Rev. B*, **2006**, 73, 235411.
49. Abramowitz, M.; Stegan, I. A.; Romer, R. H. *Handbook of Mathematical Functions with Formulas, Graphs, and Mathematical Tables* (National Bureau of Standards, Applied Mathematics Series, Washington, 1972.
50. Peres, N.M.R. Colloquium: The transport properties of graphene: An introduction. *Rev. Mod. Phys.*, **2010,** 82, No. 3, 2673.
51. Sarabalis, C. J.; Van Laer,v; Hill, J. T.; Safavi-Naeini, A. H. Optomechanical antennas for on-chip beam-steering, *Optics Express*, **2018**, 26, 17 | 20